\documentclass[doublecol,figures]{epl2} 

\newcommand{\bra}{\langle}
\newcommand{\ket}{\rangle}

\newcommand{\vecmu}{\boldsymbol{\mu}}
\newcommand{\vecxi}{\boldsymbol{\xi}}
\newcommand{\vechi}{\boldsymbol{\chi}}

\title{Plectoneme creation reduces the rotational friction of a polymer}
\shorttitle{Plectoneme creation reduces the rotational friction of a polymer} 

\author{H. Wada\inst{1} \and R. R. Netz\inst{2}}
\shortauthor{H. Wada \etal}

\institute{                    
  \inst{1} Yukawa Institute for Theoretical Physics, Kyoto University, Kyoto 606-8502, Japan\\
  \inst{2} Department of Physics, Technical University Munich, 85748 Garching, Germany
}
\pacs{87.15.H-}{Dynamics of biomolecules}
\pacs{87.16.Ka}{Filaments, microtubules, their networks, and supramolecular assemblies}
\pacs{47.15.G-}{Low-Reynolds-number (creeping) flows}

\abstract{
The torsional dynamics of a semiflexible polymer with a contour length $L$  larger
than its persistence length $L_p$ that is rotated at fixed frequency $\omega_0$ at
one end is studied by scaling arguments and hydrodynamic simulations. We find
a non-equilibrium transition at a critical frequency $\omega_\ast$:
In the linear regime, 
$\omega_0 < \omega_\ast$, axial spinning is the dominant dissipation mode.
In the non-linear regime, $\omega_0  >  \omega_\ast$, the twist-dissipation mode
involves the continuous creation of plectonemes close to the driven end
and  the rotational friction is substantially reduced.
}

\begin{document}

\maketitle

\section{Introduction}
The properties of semiflexible polymers have received continuous interest and attention 
from biophysicists for the last two decades. 
This is so since they constitute a minimal yet realistic mechanical model
for biological macromolecules such as DNA and filamentous proteins.
The framework of statistical mechanics provides various equilibrium 
properties of semiflexible polymers such as the force-extension relationship\cite{marko},
which have been successfully tested through detailed comparison with
experimental data obtained from  newly developed 
micromanipulation techniques employing optical tweezers and 
magnetic beads~\cite{ritort}.
The importance of torsional elasticity has been recognized in
the context of DNA supercoiling~\cite{marko2}.
Driven by experiments where both stretching and twist of single biopolymers could
be controlled~\cite{strick}, theories  were generalized to include twisting effects as well
and provide quantitative agreement on the static level~\cite{marko3}.

The in-vivo functioning of DNA involves non-equilibrium 
dynamic twist effects that mostly have to do with the activity of 
various DNA-processing  proteins~\cite{lavelle}.
For example, in replication, the process by which a DNA chain
is duplicated into two identical daughter strands, the helical nature
of DNA requires daughter strands to unwind and thus the mother strand to
rotate around its axis. This large scale motion has been
considered a conceptual obstacle; it was resolved by a simple
calculation demonstrating that the friction dissipation associated with
 simple axial-spinning of DNA (like in a speedometer cable) is rather small compared to 
typical biological free energies~\cite{levinthal}.
In transcription, the process by which the DNA informational content is 
copied into a continuously growing RNA chain, it was suggested that a long
nascent RNA chain might (either due to its own hydrodynamic friction 
or via anchoring to some other cellular component) provide enough 
rotational resistance to force the DNA strand to rotate 
around its own axis\cite{liu-wang}.  Experimental in-vivo and in-vitro studies found
large degrees of DNA supercoiling upon transcription, being positive in front
and negative behind the transcriptional complex\cite{liu-wang}. 
This finding, however, is at  odds with the 
above-mentioned simple axial-spinning scenario~\cite{levinthal}, since the rotational
friction of bare DNA seems not large enough to induce sufficient
torsional stress needed to induce supercoiling~\cite{nelson}.
In an effort to reconcile the conflicting experimental results, Nelson
introduced the notion of static bends along DNA,
which substantially  increase the rotational friction~\cite{nelson}.
So far, the issue of the rotational friction of a semiflexible 
chain is still a matter of interest and debate.
Single-molecule experiments on rotating DNA failed to observe
the static-bend-induced anomaly of the rotational friction, but this
could possibly be related to the  high rotational frequency 
employed in the experiments\cite{thomen}. 
Several studies did not provide a clear-cut answer as to what the 
influence of static DNA bends  is and whether other factors present in 
in-vivo studies are needed to induce substantial supercoiling of
DNA under twist injection\cite{krebs,leng,stupina,dekker}.

\begin{figure}[htb]
\begin{center}
\onefigure[width=0.999\linewidth]{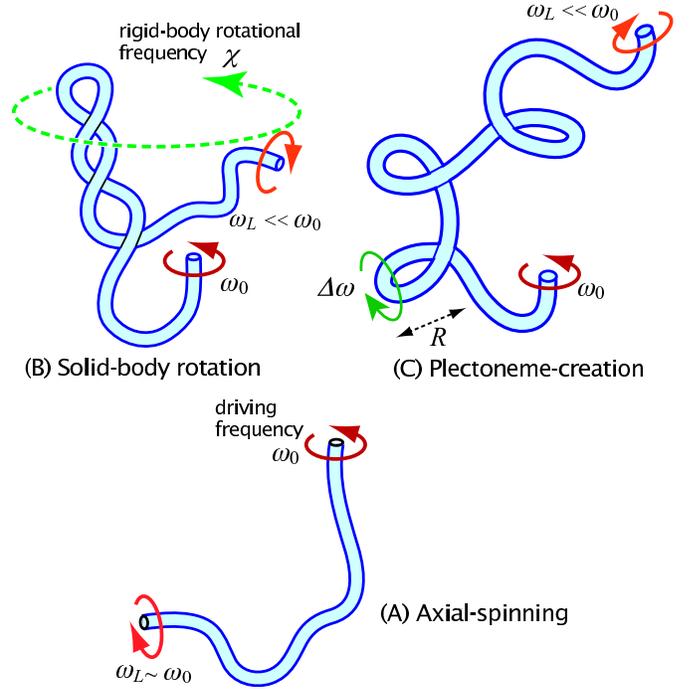}
\end{center}
\caption{A semiflexible polymer is rotated at frequency $\omega_0$ at a fixed end
and exhibits rotation at frequency $\omega_L$ at its free end.
The three  twist-dissipation channels are
(A) axial-spinning, (B) solid-body rotation, and (C) plectoneme-creation. 
}
\label{fig1}
\end{figure}

In this Letter we use non-equilibrium scaling arguments
and hydrodynamic simulations and consider a  fundamental
scenario, namely a homogeneous 
semiflexible polymer  in the stationary limit,  that is axially 
rotated at one end  at frequency $\omega_0$ with the other end free, 
as schematically depicted in Fig.~\ref{fig1}.
In the absence of shape fluctuations, i.e. at zero temperature, 
an elasto-hydrodynamic instability occurs at a
 critical driving frequency $\omega_c$ that separates a twirling
 from a whirling regime~\cite{wolgemuth};
for $\omega_0<\omega_c$ the rod stays straight and undergoes simple axial
spinning (twirling), while for $\omega_0>\omega_c$ the rod  buckles 
and displays a combination of axial spinning 
and rigid-body rotation (whirling). 
Including thermal effects, elastic shape 
fluctuations of the rod round off the transition and shift the instability  
to lower frequency~\cite{wada-netz-2006}. 
In the present study, we  consider the opposite limit
where the   chain contour length is much larger than  bend and twist
persistence lengths and
 the twirling-whirling instability is completely washed out by thermal  fluctuations.
This is relevant to transcription-driven DNA supercoiling,
but has not been studied as an elasto-hydrodynamic problem for a homogenous
semiflexible chain before. 
As our main result, we find in addition to the 
well-known axial-spinning regime,
realized for low driving frequency $\omega_0 < \omega_\ast$,  a
novel plectoneme dominated  regime at high frequencies 
$\omega_0  >  \omega_\ast$.
In this regime,  twist is converted locally into writhe (in the form
of plectonemes) close to the driven end and then diffuses out to the free end.
For sufficiently long chains, 
the crossover frequency $ \omega_\ast$ is much larger than the 
twirling-whirling threshold $\omega_c$. 
Quite surprisingly, in the plectoneme regime
the  overall rotational friction is significantly reduced as compared to the axial-spinning
scenario, i.e.,  the plectoneme  dissipation channel
transforms the injected twist into writhing motion at very low frictional cost. 
This mechanism  could be biologically relevant since it shows how 
transcription might produce
positively super-coiled DNA structures that could favorably interact with 
negatively super-coiled nucleosomal structure ahead of the RNA 
polymerase\cite{lavelle} even in topologically open systems
and at very small energy expenditure.
\section{Model}
Within linear elasticity theory, the elastic energy of an isotropic rod of contour length $L$ is given by
\begin{eqnarray}
 E_{el} &=& \int_0^L ds \left[\frac{A}{2}\kappa(s)^2+
	\frac{C}{2} \Omega(s)^2\right],
 \label{eq1}
\end{eqnarray}
where $A$ and $C$ are the bend and twist rigidities,
$\kappa(s)=|\partial^2_s{\bf r}|$ is the local curvature 
with ${\bf r}(s)$ the position of the rod centerline and $\partial_s$
denoting the derivative with respect to the arclength $s$. 
The local twist density $\Omega(s)$
is defined by $\Omega(s) = \partial_s \phi(s)$ 
where $\phi$ is the rotational angle about the local tangent $\partial_s{\bf r}$.
Note that $\Omega$ in eq.~(\ref{eq1}) 
includes also the  twist contribution due to  geometric torsion,
i.e.  changes of $\phi(s)$ are  also generated e.g.  by  out-of-plane bending~\cite{goldstein,chilico}.
%
%

In our simulations, a filament is modelled as a chain
of $N+1$ connected spheres of diameter $a$.
The total energy of the system,
$E=E_{st}+E_{el}+E_{LJ}$,
includes a stretching
contribution ensuring connectivity of spheres, 
$E_{st}=K/2\sum_{j=1}^N(|{\bf r}_{j+1}-{\bf r}_j|-a)^2$, where
${\bf r}_j$ is the position vector of monomer $j$,
and a truncated Lennard-Jones potential to account for chain
self-avoidance, 
where $E_{LJ} = \epsilon_{LJ} \sum_{i<j}[(a/r_{ij})^{12}
-2(a/r_{ij})^{6}]$ applied only for $r_{ij}=|{\bf r}_i-{\bf r}_j|<a$.
The local elastic translational force ${\bf F}_j$ and torque 
about the local tangent $T_j$ acting on each sphere are calculated
using the variational method described previously~\cite{chilico}, 
leading to the coupled Langevin equations:
$\partial_t{\bf r}_i= 
\sum_{j=1}^{N+1}\vecmu_{ij}\cdot {\bf F}_j+\vecxi_i(t)$ and 
$\partial_t \phi_i = (\pi\eta a^3)^{-1}T_i+\Xi_i(t)$,
where $\phi_i$ is the rotation angle around the bond vector
${\bf r}_{i+1}-{\bf r}_i$.
Hydrodynamic interactions between two spheres $i$ and $j$ are
included via the Rotne-Prager mobility tensor $\vecmu_{ij}$~\cite{ermak}.
For the translational self-mobility of 
the spherical monomers with diameter $a$ we use
the standard Stokes formula 
$\vecmu_{ii}={\bf 1}/(3\pi\eta a)\equiv\mu_0{\bf 1}$.
The vectorial random displacements $\vecxi(t)$ and $\Xi(t)$ model
the coupling to a heat bath and obey the fluctuation-dissipation
relations implemented numerically by a Cholesky factorization~\cite{ermak}.
For the numerical integrations we discretize the Langevin equations
with a time step $\Delta$ and rescale all lengths,
times and energies, leading to the dimensionless parameter
$\tilde{\Delta}=\Delta k_BT\mu_0/a^2$.
We set the twist-bend rigidity ratio to $C/A=1$
and the stretching modulus to $K/k_BTa^2=10^3$,
which gives negligible bond length fluctuations.
The self-crossing of the chain is entirely prevented
by setting $\epsilon_{LJ}/k_BT=10$.
For sufficient numerical accuracy we choose a time step
$\tilde{\Delta}=0.0004$.
Output values are calculated every $10^3$-$10^4$ steps, total 
simulation times are $10^{7}$ steps,
the first $10^6$ steps are not included in the data analysis.
The boundary condition at the forced end,
$\partial_s{\bf r}(0)=\hat{\bf z}$, is realized by fixing the first two monomers
in space by applying virtual forces,
which also act (via the mobility tensor) on the rest of the filament.
The rotational driving at the base
imposes $\partial_t{\phi}_1=\omega_0$, while 
force- and torque-free boundary conditions are adopted for the other end.
The number of beads is in the range  $L/a=N=30-100$.
Throughout this study, the persistence length is 
set to $L_p=10a$, thus $L/L_p=3-10$.

\section{Buckling and plectonemes}
\begin{figure}
\begin{center}
\onefigure[width=0.7\linewidth]{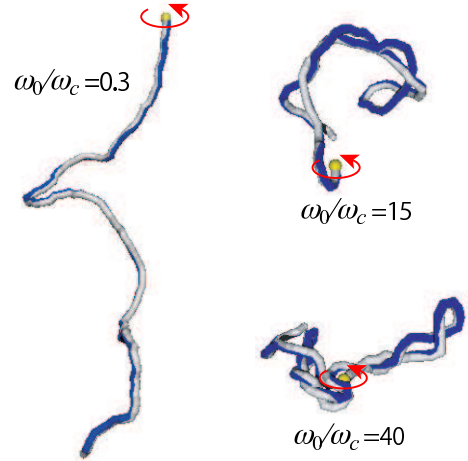}
\end{center}
\caption{Typical chain conformations from our hydrodynamic simulations for 
chain length $L=100a$ and persistence length $L_p=10a$ and rotational frequency
$\omega_0/\omega_c=0.3, 15$ and 40.}
\label{fig2}
\end{figure}

To set the stage, we first repeat the zero-temperature
scaling argument for the critical frequency $\omega_c$
at which an axially rotated rod exhibits the buckling instability.
At low rotational frequency, $\omega_0 < \omega_c$, 
the rod is twisted but remains straight and 
the torque at the base balances the total rotational drag,
$\pi\eta a^2\omega_0 L \sim C\Omega(0)$.
On the scaling level,
the rod buckles when the twisting torque, $C\Omega(0)$, becomes 
comparable to the bending torque, $A/L$, 
giving the  critical frequency 
$\omega_c \sim A/(\pi\eta a^2L^2)$ independent of the twist 
rigidity $C$~\cite{wolgemuth}.
The linear stability analysis predicts 
an instability at $  \omega_c  \equiv  8.9 A/(\pi\eta a^2L^2) $ ~\cite{wolgemuth},
the numerical prefactor was confirmed  by 
simulations in the zero-temperature-limit, i.e. for 
 a very stiff polymer with  persistence length 
$L_p = A/k_BT \gg L$~\cite{wada-netz-2006}.
In Fig.~\ref{fig2}  we show typical chain snapshots obtained in our
dynamic simulations  for driving frequencies $\omega_0/\omega_c=0.3$, 15 and 40
for a rather flexible chain with $L/L_p = 10$.
For $\omega_0<\omega_c$, the chain flexes randomly due
to thermal motions and spins about its local axis 
at frequency $\sim \omega_0$.
For $\omega_0 \gg \omega_c$, in contrast,
 the polymer exhibits  continuous generation of plectoneme-like
 structures and their  diffusive transport from the driving end to the free end.
In the following we  present a much-simplified scaling theory,
valid for long elastic rods  $L \gg L_p$,
that establishes  a minimal framework for treating
the competitive twist transport due to 
axial spinning and plectoneme creation.

\section{Scaling}
There are three different ways for the polymer to transport the injected twist
 to its free end, see fig.~\ref{fig1}.
The first one is the (A) axial spinning mode, where the 
polymer rotates around its contour like a speedometer cable,
 which we show to be the dominant dissipation mode for low enough
driving frequencies. 
The second one is the  
(B) solid-body rotation mode, where the whole polymer coil
whirls around the rotational axis at some frequency $\chi$.
The third one is the (C) plectoneme creation/diffusion mode, in which plectoneme-like structures are 
continuously generated at the rotated end  and are transported diffusively towards
the free end. As we show in this paper, 
this dynamic twist-writhe conversion is a highly nonlinear mechanism 
that provides a very efficient means of relieving  torsional stress at elevated driving frequencies.
\begin{figure}
\begin{center}
\onefigure[width=0.999\linewidth]{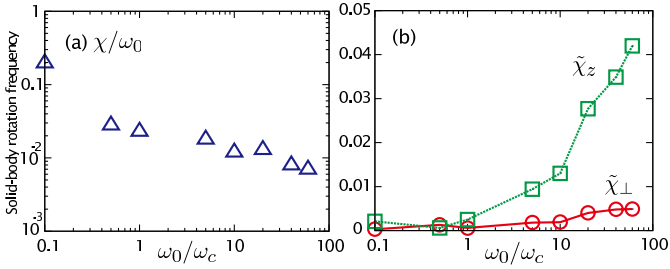}
\end{center}
\caption{(a) Relative solid-body rotation frequency, $\chi/\omega_0$, as a function of 
$\omega_0/\omega_c$. (b) Rescaled components of the vector $\vechi$, 
$\tilde{\chi}_{\perp}=\chi_{\perp}a^2/(\mu_0k_BT)$,
and $\tilde{\chi}_z=\chi_za^2/(\mu_0k_BT)$, as a function of $\omega_0/\omega_c$. 
Note that $\chi=(\chi_{\perp}^2+\chi_z^2)^{1/2}$.
The data are obtained for  $L=50a$ and $L_p=10a$.}
\label{fig3}
\end{figure}
Noting that in a stationary state, the twist that is injected into the polymer, $\omega_0$, 
has to exit the chain at the
free end either in the form of axial spinning, solid-body rotation of frequency $\chi$, or writhe, we write 
\begin{eqnarray}
 \omega_0 & = & \omega_L+\chi + \Delta \omega,
 \label{eq:new1}
\end{eqnarray}
where $\omega_L$ is the axial spinning frequency of the free end 
and $\Delta \omega$ denotes the fraction that is converted into writhe.
In order to decide which of these three channels is in fact realized, we use
the concept of minimum entropy production for non-equilibrium stationary states~\cite{mazur},
according to which the state of least dissipation is stable. We will later confirm each of our
scaling results by our simulations, which gives further credibility to our scaling approach.
We thus have to estimate the power dissipation in each of the modes depicted in fig.~\ref{fig1}.
The power dissipation due to  axial spinning  is 
 $P_{as}\sim \eta La^2\omega_L^2$. Here we assume that average axial spinning frequency
 is $\omega_L$, meaning that in the plectoneme regime the rotational profile 
$\omega(s)$ decays very quickly
(in fact exponentially) along the chain contour to the value  $\sim \omega_L$.
This is confirmed by our simulations,  see fig.~\ref{fig4} (b).
Likewise, the power dissipation due to solid-body rotation is given as 
$P_{sb}\sim \eta R^3\chi^2$. Except for a compact globule with $R^3 \sim a^2 L$
we see that axial spinning is a less costly channel for twist transport
than solid-body rotation, i.e. $P_{as} \ll P_{sb}$.
We therefore neglect solid-body rotation in what follows.
The plectoneme-creation channel is more complicated.
We consider the creation of a single loop of radius $R$  nearby the driven end,
see fig.~\ref{fig1} (C).
Extension to a more complex  plectoneme structure involving multi-loops  
(as actually observed in the simulations, see fig.~\ref{fig2}) 
is straightforward, but does not change the conclusions on the scaling level.
The bending energy of  one loop  is $k_BT L_p/R$, the loop creation frequency
is $\Delta \omega$,  thus the
power consumption is  $P_{loop}\sim k_BT  \Delta \omega   L_p/R$.
In order to generate a loop of radius  $R$ nearby the driven end, the whole chain has to 
slide by the excess length  $2\pi R$  during the time scale of $(\Delta \omega)^{-1}$.
The frictional force associated with  this sliding with velocity $\sim R \Delta \omega$  is
$F_{slide} \sim \eta L R \Delta \omega /\ln(L/a)$.
Neglecting the  logarithmic hydrodynamic correction,
the power dissipation of sliding is 
$P_{slide} \sim F_{slide} R   \Delta \omega  \sim \eta L( R \Delta \omega)^2$.
Minimizing the total plectoneme creation dissipation,  
 $P_{plec}=P_{loop}+P_{slide}$,  with respect to the loop radius $R$, we obtain
\begin{eqnarray}
 P_{plec} &\sim& (k_BTL_p \Delta \omega )^{4/3}(\eta L)^{1/3},
 \label{eq:new2}
\end{eqnarray}
 with the loop size given by $R^3\sim k_BTL_p/(\eta L \Delta \omega)$.
 Due to the fractional power law $ P_{plec} \sim (\Delta \omega )^{4/3}$, it is easy
 to see that plectoneme creation is unfavorable at low frequencies but will win 
 over axial spinning at high frequencies.
 \begin{figure}
\begin{center}
\onefigure[width=0.70\linewidth]{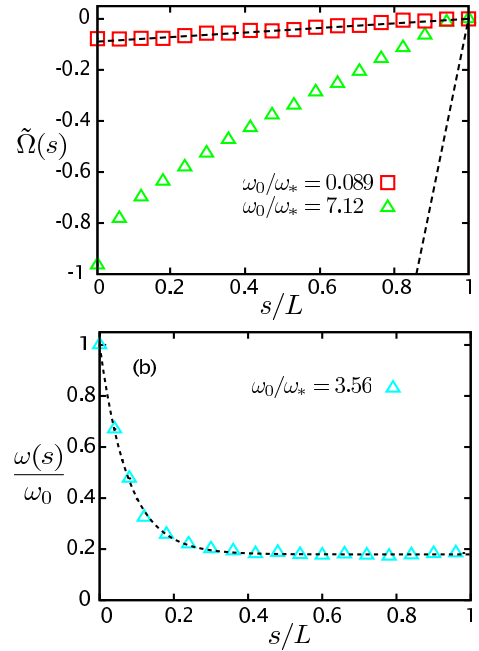}
\end{center}
\caption{(a) Steady-state profile of the rescaled twist density, $\tilde{\Omega}(s)=\Omega(s)a$, for two different
driving frequencies, $\omega_0/\omega_{\ast}=0.089$ (squares) and 7.12 (triangles). 
The broken lines are the prediction from  linearized theory, see text.
(b) Steady-state profile of the rescaled rotational velocity, $\omega(s)/\omega_0$
together with an exponential fit (broken line), for $\omega_0/\omega_{\ast}=3.56$.
All data were obtained for $L=50a$ and $L_p=10a$. }
\label{fig4}
\end{figure}
We now minimize the total dissipation  $P \sim P_{as}+P_{plec} $
with respect to the unknown frequency at the free end, $\omega_L$, and use
eq.~(\ref{eq:new1}) and $\chi=0$. For low frequency  we obtain pure axial spinning,  
$\omega_L \sim \omega_0$, i.e. the twist that is injected at one end comes
out as axial spinning at the other end.
For high frequencies $\omega_0 > \omega_{\ast}$,  with a crossover frequency
defined as 
\begin{equation} \label{wast}
\omega_{\ast} \equiv  k_BTL_p/(\pi \eta a^3L) \sim  \omega_c(L/a)  ,
\end{equation}
 on the other hand, 
we obtain
 \begin{eqnarray}
 \omega_L &\sim&   \omega^{2/3}_{\ast} \omega_0^{1/3},
 \label{eq:new3a}\\
 P &\sim&  \eta L a^2 \omega^{2/3}_{\ast}  \omega_0^{4/3}.
  \label{eq:new3b}
\end{eqnarray}
 Interestingly, the cross-over frequency $  \omega_{\ast}$ is much larger than the zero-temperature
twirling-whirling critical frequency $\omega_c$ by a factor proportional to the chain contour length, $L/a$,
and the dissipation $P$  is reduced compared to the axial spinning scenario $P_{as}$.

To confirm these predictions, we extract the solid-body rotation rate vector, $\vechi$, 
from our hydrodynamic simulations.
For rigid-body rotation, i.e, 
$\partial_t{\bf r}\simeq \vechi\times{\bf r}$, we obtain
the vector  $\vechi=(\chi_{\perp},\chi_z)$ via
$\vechi={\bf I}^{-1}\cdot{\bf L}$, where ${\bf I}=\sum_{j=1}^N[{\bf r}_j{\bf 1}-{\bf r}_j{\bf r}_j]$ is the moment of
inertia tensor and ${\bf L}=\sum_{j=1}^N{\bf r}\times d{\bf r}/dt$ is the polymer angular momentum 
(the mass of the polymer beads is set to unity).
Fig.~\ref{fig3}b)  shows  $\vechi=(\chi_{\perp},\chi_z)$  as a function of the rescaled driving frequency, $\omega_0/\omega_c$;
solid-body rotation is small
 for $\omega_0<\omega_c$, while $\chi_z$ 
grows significantly beyond $\omega_c$.  The hydrodynamic shear due to this rotation deforms
the plectonemes at high values of $\omega_0$, which is a secondary effect that 
we neglect in the current version of our scaling arguments.
In fig.~\ref{fig3} (a), we show that $\chi/\omega_0=|\vechi|/\omega_0$ is much smaller than unity,
which confirms that  solid-body rotation is negligible compared to other dissipation channels.

The total twist $Tw$ is defined as the integrated twist density
along the chain arclength, $Tw=(2\pi)^{-1}\int_0^Lds \Omega(s)$.
In the axial-spinning regime, i.e., $\omega_0<\omega_{\ast}$,  chain shape fluctuations are
decoupled from twisting motion, and the injected twist propagates diffusively 
along the chain~\cite{nelson}.
The twist $\Omega$ thus obeys the linear diffusion equation, $\pi \eta a^2\partial_t\Omega=C\partial_s^2\Omega$.
In steady state under the boundary condition, $C\Omega(0)=\pi\eta a^2\omega_0L$,
we thus obtain $\Omega(s)  =  (\pi\eta a^2\omega_0/C) (s-L)$. 
This is in quantitative  agreement with simulation for $\omega_0/\omega_{\ast}=0.089$,
as seen in fig.~\ref{fig4} (a) (open squares and broken line). 
The total twist in this regime is given by
$ Tw \simeq -\pi \eta a^2\omega_0L^2/(4\pi C)=-(A/4\pi C)(L/a)(\omega_0/\omega_{\ast})$
and thus is  linear in the driving frequency $\omega_0$.
In the plectoneme regime, on the other hand, the twist is much reduced compared to the linear
law (lower broken line in fig.~\ref{fig4} (a)) and shows a nonlinear spatial profile (open triangles in fig.~\ref{fig4} (a)).
In this regime, the twist density $\Omega$ receives an additional contribution from the
geometric torsion due to plectoneme formation, which however we  neglect.
Assuming the twist to be proportional to the average spinning frequency $\omega_L$ in the region
where plectonemes have already formed, i.e.
$Tw \sim (AL/aC)(\omega_L/\omega_{\ast}) \sim (\omega_0/\omega_{\ast})^{1/3}$,  we thus predict
\begin{eqnarray}
 Tw &\sim&
 \left\{
 \begin{array}{ll}
 \displaystyle{\frac{A L }{4\pi a C}\frac{\omega_0}{\omega_{\ast}}} & (\omega_0<\omega_{\ast})\\
  & \\
 \displaystyle{\frac{AL} {a C}\left(\frac{\omega_0}{\omega_{\ast}}\right)^{1/3}} & (\omega_0>\omega_{\ast}).\\
 \end{array}
 \right.
 \label{eq:scale-twist}
\end{eqnarray}
Finally, the effective rotational friction coefficient $\Gamma_r$  is defined via 
$P\sim \Gamma_r \omega_0^2$, where $P$ is the total dissipation, given  by eq.~(\ref{eq:new3b})
in the plectoneme regime.  We  obtain
\begin{eqnarray}
 \frac{\Gamma_r}{\pi \eta a^2L} &\sim&
 \left\{
 \begin{array}{ll}
 1 & (\omega_0<\omega_{\ast})\\
  & \\
 \displaystyle{\left(\frac{\omega_0}{\omega_{\ast}}\right)^{-2/3}} & (\omega_0>\omega_{\ast}).\\
 \end{array}
 \right.
  \label{eq:scale-gamma}
\end{eqnarray}
For $\omega_0<\omega_{\ast}$,  the rotational friction  corresponds to the axial spinning scenario,
while for  $\omega_0 > \omega_*$ a drastic decrease is predicted.
Note that  for  $\omega_0 > \omega_*$  the dependence on chain length 
is $\Gamma_r \sim  \omega_0^{-2/3} L^{1/3}$ and thus increases only sub-linearly with polymer length.

\begin{figure}
\begin{center}
\onefigure[width=0.85\linewidth]{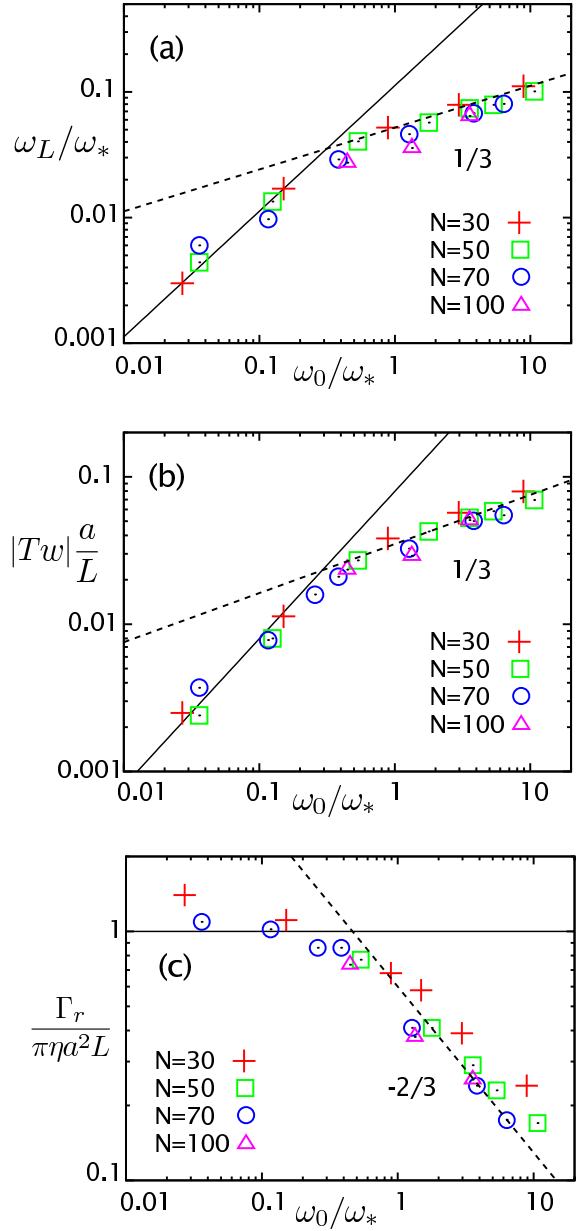}
\end{center}
\vspace*{-0.75cm}
\caption{Scaling plots of hydrodynamic simulation results.
(a) Rescaled axial spinning frequency at free end, $\omega_L/\omega_{\ast}$, 
(b) total twist divided by the chain length, $|Tw | (a/L))$, 
and (c) rescaled rotational friction constant, $\Gamma_r/(\pi \eta a^2 L)$,
plotted as a function of $\omega_0/\omega_{\ast}$.
Soild lines denote  the predictions from linear theory in the axial spinning regime valid  
for $\omega_0<\omega_\ast$, broken lines denote 
the non-linear results valid in the plectoneme regime as given 
 in eqs.~(\ref{eq:new3a}), (\ref{eq:scale-twist}) and (\ref{eq:scale-gamma}).
The crossover between the regimes occurs around a value of $\omega_0 = c_\ast \omega_\ast$
with $ c_\ast = 0.4 \pm 0.1$.
}
\label{fig5}
\end{figure}

In Fig.~\ref{fig5},  $\omega_L$, $Tw$ and $\Gamma_r$ from simulations are
plotted as a function of $\omega_0/\omega_{\ast}$
and confirm the  analytical predictions for the non-linear plectoneme regime ,
eqs.~(\ref{eq:new3a}), (\ref{eq:scale-twist}) and 
(\ref{eq:scale-gamma}).
In the simulations, the torque $M_0$ at a given driving frequency $\omega_0$ is
measured  and averaged to obtain $\Gamma_r=\bra M_0\ket/\omega_0$.
In particular, the crossover frequency between the axial-spinning and the plectoneme
regime is quite consistently found to occur at 
$\omega_0=  c_\ast  \omega_\ast  \sim (L/a) \omega_c$
with $c_\ast = 0.4 \pm 0.1$ and $\omega_\ast $ defined in eq. (\ref{wast}). 
Note that the whirling instability, at $\omega_0 \simeq \omega_c$ and realized only
for $L/L_p < 1$, is conceptually distinct from the plectoneme transition at $\omega_0 \simeq \omega_\ast$,
which is only observable for long chains $L/L_p \gg 1$. In fact, the two transitions do not merge
or interconnect at intermediate values of $L/L_p$:  for  $L/L_p \simeq 1$ a
semiflexible chain  rather shows a continuous shape and rotational mode evolution 
with increasing $\omega_0$ without a sharply defined transition.

\section{Discussion}
For a hydrodynamic diameter of ds-DNA of $a\sim 2$ nm,
a bend persistence length $L_p\sim 30$ nm and length $L\sim 12 \mu$m, 
we obtain  according to eq. (\ref{wast})
a cross over frequency $c_\ast \omega_{\ast} \sim 2  \times10^5$ rad/s.
The rotational friction  of DNA molecules of length $L\sim 12$ $\mu$m has 
been measured  for rotational frequencies up to $\omega_0\sim 12000$ rad/s in 
DNA unzipping experiments~\cite{thomen} and showed no detectable non-linear 
frequency dependence, in agreement with our estimate for the threshold $\omega_\ast$.
On the other hand, for longer chains or
in a crowded cellular environment with a much elevated viscosity,
 the crossover frequency $\omega_{\ast}$   can be lowered to an 
 experimentally  reachable value.
The critical torque  $M_\ast  =  \pi\eta a^2L c_\ast \omega_{\ast} 
\simeq  c_\ast  \pi k_BT L_p/a$ is independent of both viscosity
and chain length and for bare DNA is of the order of a few tens of  $k_BT$,  too high 
for, e.g.,  single {\it Escherichia coli} RNA polymerase to achieve.
But for chromatin structures, the effective bending persistence length 
$L_p^{chrom}$  has been shown
to stay rather constant while the effective radius $a^{chrom}$ is increased
substantially\cite{lavelle}, 
meaning that the critical torque $M_\ast^{chrom}$  for a chromatin fiber
might be of the  order of $k_BT$, quite in reach of 
typical torques generated by polymerase.

In recent experiments,
the elongation dynamics of supercoiled DNA in response to sudden increase of tension 
was studied~\cite{dekker}.
The dynamics showed almost no effect of the continuous removal
of plectonemes, setting an upper bound on the hydrodynamic
friction associated with DNA rotation, which turns out to be in agreement 
with the axial spinning scenario and thus is  consistent with our  results.
Previous {\it in vitro} and {\it in vivo} studies
on transcriptionally-driven
DNA supercoiling have been interpreted as evidence of enhanced rotational 
friction when compared to the simple axial-spinning estimate~\cite{liu-wang}. 
Krebs and Dunaway showed  {\it in vivo} 
that polymerase  drives DNA supercoiling during the transcription process 
for   linear DNA templates 
longer than 17 to 19 kbp~\cite{krebs}. 
One explanation involves  static DNA  bends  
or kinks stabilized by DNA-binding proteins, which are suggested to significantly
slow down the torsional relaxation and lead to a considerably
enhanced rotational drag torque~\cite{nelson,stupina,leng}.
It remains to study how static DNA bends interfere with the 
plectoneme creation/diffusion 
scenario developed in this paper.

To summarize, using scaling arguments and hydrodynamic simulations, we have studied
the stationary nonequilibrium dynamics of torsionally driven semiflexible polymers.
For chains much longer than their persistence length, $L>L_p$, two dynamical regimes are distinguished:
for small driving frequency, $\omega_0 < \omega_\ast$,
we find the standard  axial-spinning regime 
where the chain flexes randomly and spins about its local axis.
For  large rotational frequency, $\omega_0  >  \omega_\ast$,
twist is locally converted into writhe
close to the driven end and then diffuses out to the free end without much concerted solid-body rotation.
In this  plectoneme  regime the filament  exhibits only minimal axial spinning
and a significant reduction of the rotational friction as compared to axial spinning is obtained.
Two further conclusions might be biologically relevant:
The nature of the twist-writhe conversion process leads to a narrow spatial localization
of the twist density close to the rotated part of the chain, 
which in-vivo might guide and concentrate the activity
 of twist-sensitive proteins to a region close to the polymerase complex.
Finally, the positive supercoiling, created in front of the point of twist-injection
in the form of plectonemes, would be a simple
physical mechanism for loosening or even driving off histones 
from their nucleosomale core particles, the negative supercoiling behind 
the twist-injection point would form a template for strengthening or reforming
the nucleosomal core structure.

\acknowledgments
We thank C. Lavelle and P. Nelson for useful comments.
Financial support from MEXT-Japan (Grant in Aid, No.20740241)
and the Excellence Cluster Nano-Initiative-Munich is acknowledged. 


\begin{thebibliography}{99}

\bibitem{marko}
 \Name{Vologodskii A}
  \REVIEW{Macromolecules}{27}{1994}{5623};
 \Name{Marko J. F., \and Siggia E. D.}
  \REVIEW{Macromolecules}{28}{1995}{8759};
 \Name{Hallatschek O., Frey E. \and Kroy K.}
  \REVIEW{Phys. Rev. E}{75}{2007}{031905}.


\bibitem{ritort}
 \Name{Ritort F}
  \REVIEW{J. Phys.: Condens. Matter.}{18}{2006}{R531} and references therein.


\bibitem{marko2}
 \Name{Marko J. F. \and Siggia E. D.}
  \REVIEW{Phys. Rev. E}{52}{1995}{2912}.


\bibitem{strick}
  \Name{Strick T. R., Allemand J.-F., Bensimon D., Bensimon A. \and 
 Croquette V.}
  \REVIEW{Science}{271}{1996}{1835}.


\bibitem{marko3}
 \Name{Marko J. F.}
  \REVIEW{Phys. Rev. E}{55}{1997}{1758};
 \Name{Vologodskii A. V. \and Marko J.F.}
  \REVIEW{Biophys. J.}{73}{1997}{123};
 \Name{Moroz J. D. \and Nelson P.}
  \REVIEW{Proc. Nat. Acad. Sci. USA}{94}{1997}{14418}.
 
\bibitem{lavelle}
\Name{Lavelle C.}
  \REVIEW{Biochem. Cell. Biol.}{87}{2009}{307}.
 


\bibitem{levinthal}
 \Name{Levinthal C. \and Crane H.} 
 \REVIEW{Proc. Natl. Acad. Sci. U.S.A.}{42}{1956}{436}.
 
 \bibitem{liu-wang}
 \Name{Liu L. F. \and Wang J. D.}
  \REVIEW{Proc. Natl. Acad. Sci. U.S.A.}{84}{1987}{7024};
 \Name{Tsao Y.-P., Wu H.-Y. \and Liu L. F.}
  \REVIEW{Cell}{56}{1989}{111};
\Name{Droge P. \and Nordheim A.}
  \REVIEW{Nucleic Acids Research}{19}{1991}{2941}.

\bibitem{nelson}
 \Name{Nelson P.}
  \REVIEW{Proc. Natl. Acad. Sci. U.S.A.}{96}{1999}{14342}.
  
\bibitem{thomen}
 \Name{Thomen P., Bockelmann U. \and Heslot F.}
 \REVIEW{Phys. Rev. Lett.}{88}{2002}{248102};
 \Name{Nelson P.}
 \REVIEW{Phys. Rev. Lett.}{92}{2004}{159801};
 \Name{Thomen P. \and Heslot F.}
 \REVIEW{Phys. Rev. Lett.}{92}{2004}{159802}.

\bibitem{krebs}
 \Name{Krebs J. E. \and Dunaway M.}
 \REVIEW{Mol. Cell. Biol.}{16}{1996}{5821}.

\bibitem{leng}
 \Name{Leng F. \and McMacken R.}
 \REVIEW{Proc. Natl. Acad. Sci. U.S.A.}{99}{2002}{9139};
 \Name{Leng F., Amado L. \and McMacken R.}
 \REVIEW{J. Biol. Chem.}{279}{2004}{47564}.

\bibitem{stupina}
 \Name{Stupina V. A. \and Wang J. C.}
  \REVIEW{Proc. Natl. Acad. Sci. U.S.A}{101}{2004}{8608}. 

\bibitem{dekker}
 \Name{Crut A., Koster D. A., Seidel R., Wiggins C. H. \and Dekker N. H.}
 \REVIEW{Proc. Natl. Acad. Sci. U.S.A.}{104}{2007}{11957}.

\bibitem{wolgemuth}
  \Name{Wolgemuth C. W., Powers T. R. \and Goldstein R. E.}
  \REVIEW{Phys. Rev. Lett.}{84}{2000}{1623}.

\bibitem{wada-netz-2006}
 \Name{Wada H. \and Netz R. R.}
 \REVIEW{Europhys. Lett.}{75}{2006}{645}.

\bibitem{goldstein}
 \Name{Kamien R. D.}
 \REVIEW{Eur. Phys. J. B}{1}{1998}{1};
 \Name{Goldstein R. E., Powers T. R. \and Wiggins C. H.}
 \REVIEW{Phys. Rev. Lett.}{80}{1998}{5232}.


 
\bibitem{chilico}
 \Name{Chirico G. \and Langowski J.}
 \REVIEW{Biopolymers}{34}{1994}{415}.


\bibitem{ermak}
 \Name{Ermak D. L. \and McCammon J. A.}
 \REVIEW{J. Chem. Phys.}{69}{1978}{1352}.

\bibitem{mazur}
 \Name{de Groot S. R. \and Mazur P.}
 \Book{Non-equilibrium Thermodynamics} 
 \Publ{North-Holland Publishing Co., Amsterdam} 
\Year{1962},
see Chap. V.




\end{thebibliography}
\end{document}